\documentclass{elsart1p}
\usepackage{graphicx}
\usepackage{amssymb}
\begin{document}
\begin{frontmatter}
% use the corauthref command within \author for corresponding author
% footnotes;
\title{Update of the Unitarity Triangle Analysis}
\author{{Cecilia Tarantino\thanksref{label1} (on behalf of the UTfit Collaboration)}}
\thanks[label1]{I would like to thank the organizers of PANIC'08 for the very pleasant conference realized in Eilat.}
\address{Dip. di Fisica, Universit{\`a} di Roma Tre, and INFN Sez. di Roma Tre,\\Via della Vasca Navale 84, I-00146 Roma, Italy}
\begin{abstract}
We present the update of the Unitarity Triangle (UT) analysis within the Standard Model (SM) and beyond.
Within the SM, combining the direct measurements on sides and angles, the UT turns out to be overconstraint in a consistent way, showing that the CKM matrix is the dominant source of flavour mixing and CP-violation and that New Physics (NP) effects can appear at most as small corrections to the CKM picture.
Generalizing the UT analysis to investigate NP effects, constraints on $b \to s$ transitions are also included and both CKM and NP parameters are fitted simultaneously. While no evidence of NP effects is found in $K$-$\bar K$ and $B_d$-$\bar B_d$ mixing, in the $B_s$-$\bar B_s$ mixing an hint of NP is found at the $2.9 \sigma$ level.
The UT analysis beyond the SM also allows us to derive bounds on the coefficients of the most general $\Delta F=2$ effective Hamiltonian, that can be translated into bounds on the NP scale.

\end{abstract}
\end{frontmatter}
%
% main text
\section{Unitarity Triangle Analysis within the Standard Model}
\label{SM}
\begin{figure}[t!]
\begin{center}
\vspace{-1.0cm}
\includegraphics[scale=0.3,angle=0]{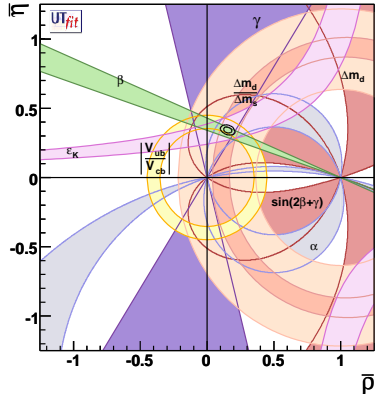} \\
\end{center}
%\vspace{-0.3cm}
\caption{\sl Constraints on the $\bar \rho$-$\bar \eta$ plane, including both angle and side measurements. The closed contours at $68$\% and $95$\% probability are shown. The full lines correspond to $95$\% probability regions for the different constraints.}
%\vspace{-0.2cm}
\label{fig:SM_allconstr}
\end{figure}

We present in this section the update of the Unitarity Triangle (UT) analysis within the Standard Model (SM), performed by the UTfit collaboration following the method described in refs.~\cite{Ciuchini:2000de,Bona:2005vz}. The constraints used in the analysis can be distinguished in side and angle constraints, where the latter do not rely on theoretical calculations of hadronic matrix elements. The side constraints come from the measurement of direct CP-violation
in the kaon sector ($\epsilon_K$), of $B_d$ and $B_s$ mixing ($\Delta m_d$, $|\Delta m_s/\Delta m_d|$) and of semileptonic B decays ($|V_{ub}/V_{cb}|$). The angle constraints are CP-violating measurements for the $B_d$-system, performed with high statistics at B-factories: $\sin 2\beta$, $\alpha$, $\gamma$, $\cos 2 \beta$, and $2 \beta + \gamma$.
As shown in fig.~\ref{fig:SM_allconstr}, the CKM matrix turns out to be consistently overconstraint.
The CKM parameters $\bar \rho$ and $\bar \eta$ result to be accurately determined: $\bar \rho=0.155\pm0.022$, $\bar \eta=0.342\pm0.014$~\cite{UTfitwebpage}.\footnote{The CKMfitter collaboration founds compatible results, though with larger uncertainties: 
$\bar \rho=0.214^{+0.031}_{-0.104}$, $\bar \eta=0.308^{+0.061}_{-0.025}$~\cite{Deschamps}.}
The UT analysis has thus established that the CKM matrix is the dominant source of flavour mixing and CP-violation and that New Physics (NP) effects can at most represent a small correction to this picture.
%%%%%%%%%%%%%%%%%%%%%%%%%%%%%%%%%%%%%%%%%%%%%%%%%%%%%%%%%%%%%%%%%%%%%
%%%%%%%%%%%%%%%%%%%%%%%%%%%%%%%%%%%%%%%%%%%%%%%%%%%%%%%%%%%%%%%%%%%%%

\begin{figure}[!h]
\vspace{-1.0cm}
\begin{minipage}{16pc}
\includegraphics[scale=0.3,angle=0]{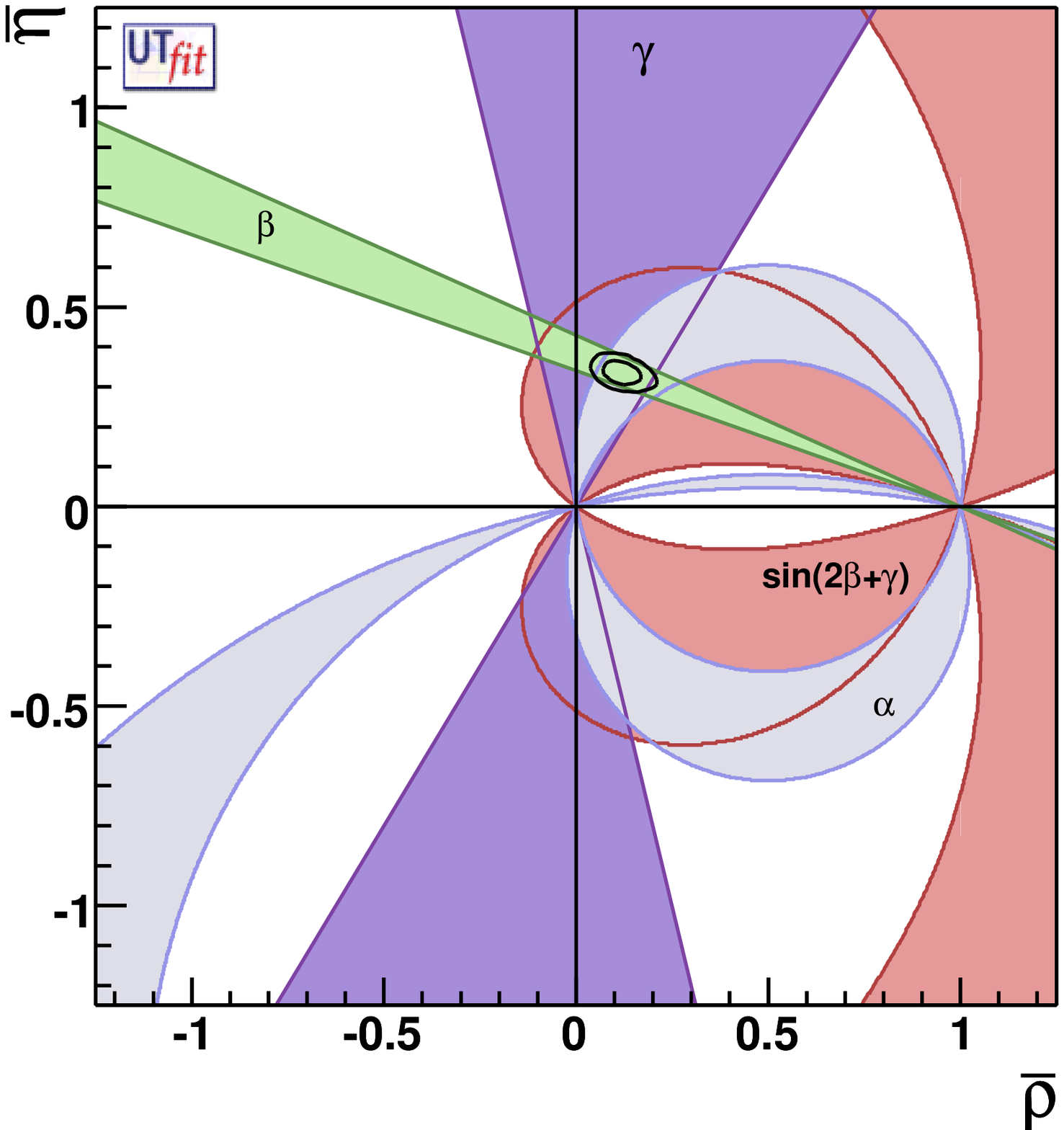}
\end{minipage}\hspace{2pc}
\begin{minipage}{16pc}
\includegraphics[scale=0.3,angle=0]{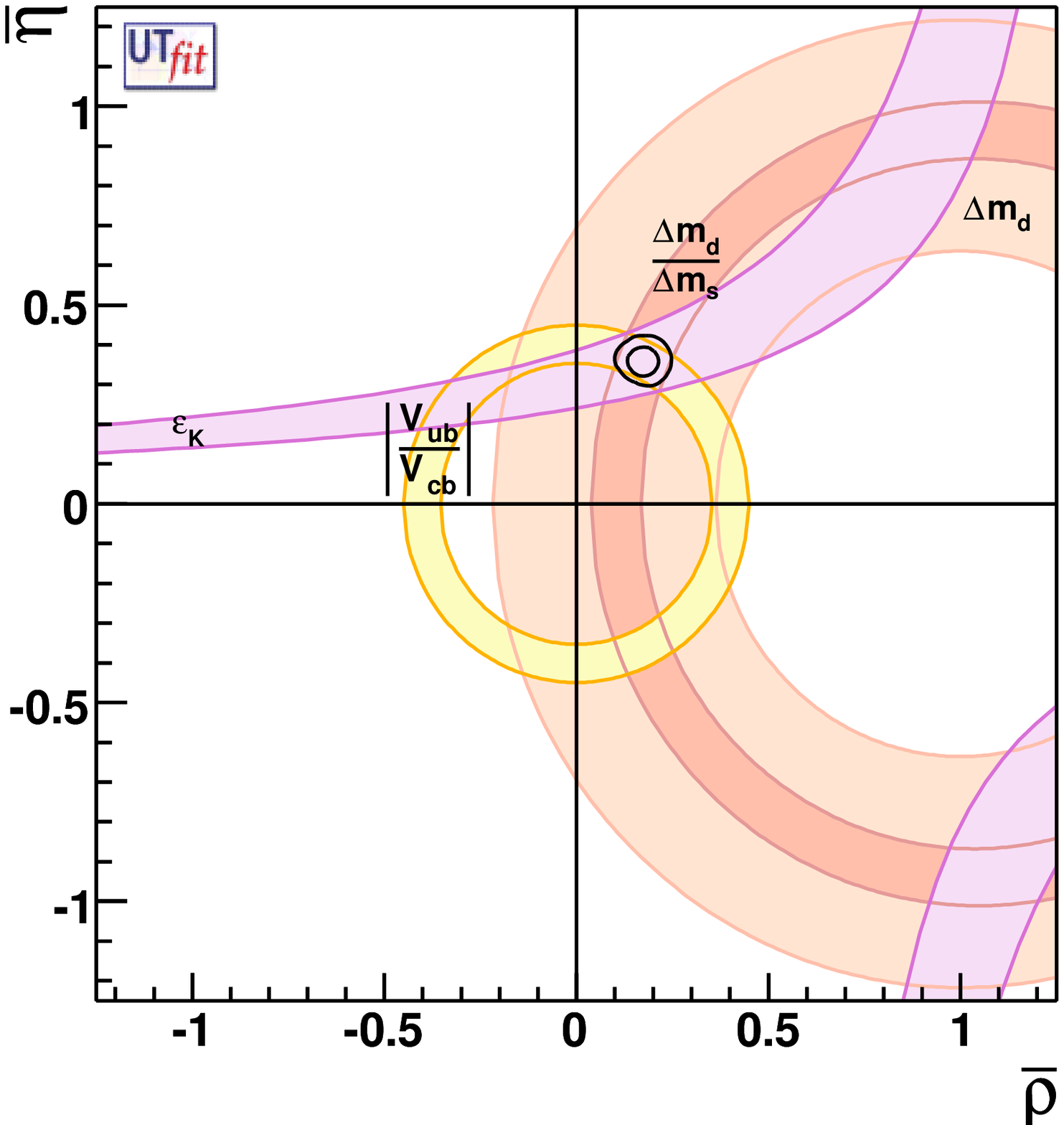}
\end{minipage}
\caption{\sl Constraints on the $\bar \rho$-$\bar \eta$ plane, including only angle (left) or side (right) measurements. The closed contours at $68$\% and $95$\% probability are shown. The full lines correspond to $95$\% probability regions for the different constraints.}
\label{fig:SM_anglevsside}
\end{figure}

Due to the redundant experimental constraints, interesting consistency checks can be performed by comparing various UT analyses where different constraints are used.
In particular, the UT analyses based on only angle (UTangle) or only side (UTside) constraints, shown in fig.~\ref{fig:SM_anglevsside}, provide well compatible results~\cite{UTfitwebpage}: $\bar \rho=0.120\pm0.034$, $\bar \eta=0.335\pm0.020$ and $\bar \rho=0.175\pm0.027$, $\bar \eta=0.360\pm0.023$, respectively.
The $\sim 1.3 \sigma$ difference between the two $\bar \rho$ results is mainly a manifestation of the tension of the $|V_{ub}|$ inclusive measurement, based on heavy quark effective theory parameters extracted from experimental fits with some model dependence, with the rest of the fit and with the $|V_{ub}|$ exclusive measurement, relying on semileptonic form factors determined from lattice QCD or QCD sum rules.
In fact, the UTangle analysis turns out provide an indirect determination of $|V_{ub}|$ ($|V_{ub}|=(34.1\pm1.8)\cdot 10^{-4}$) that is in perfect agreement with the $|V_{ub}|$ exclusive measurement ($|V_{ub}|=(35 \pm 4)\cdot 10^{-4}$), while the UTside analysis uses in input the inclusive-exclusive average for $|V_{ub}|$ that is $\sim 1.2 \sigma$ higher than the UTangle indirect determination~\cite{UTfitwebpage}.
%%%%%%%%%%%%%%%%%%%%%%%%%%%%%%%%%%%%%%%%%%%%%%%%%%%%%%%%%%%%%%%%%%%%%
%%%%%%%%%%%%%%%%%%%%%%%%%%%%%%%%%%%%%%%%%%%%%%%%%%%%%%%%%%%%%%%%%%%%%

The (overconstraint) UT analysis also allows to extract some hadronic quantities that can be compared to the results of lattice QCD calculations~\cite{Lubicz:2008am}.
This comparison is shown in Table~\ref{tab:lattice} for the hadronic parameters describing mixing in the $K$-, $B_d$- and $B_s$-meson sectors. The remarkable agreement between the lattice calculations and the indirect UT analysis determinations provides additional evidence of the SM success in describing flavour physics and of the reliability of lattice QCD calculations.
It is interesting to note that an improvement of the accuracy of the lattice determinations of $\hat B_K$ and $\xi$ would be important to increase the precision of the UT analysis.
%%%%%%%%%%%%%%%%%%%%%%%%%%%%%%%%%%%%%%%%%%%%%%%%%%%%%%%%%%%%%%%%%%
\begin{table}[h!]                                            
\begin{center}                                               
\begin{tabular}{||c||c|c|c||}                            
\hline                                                       
 & $\hat B_K$ & $f_{B_s} \sqrt{\hat B_{B_s}}$ & $\xi$ \\ \hline 
UTA & $0.75 \pm 0.07$ & $264.7 \pm 3.6$ & $1.26 \pm 0.05$ \\
Lattice & $0.75 \pm 0.07$ & $270 \pm 30$ & $1.21 \pm 0.04$ \\
\hline
\end{tabular}
\end{center}
%\vspace{-0.2cm}
\caption{\sl Values of the hadronic parameters that describe $K$-$\bar K$ and $B_{d,s}$-$\bar B_{d,s}$ mixing: $\hat B_K$, $f_{B_s} \sqrt{\hat B_{B_s}}$ and $\xi \equiv (f_{B_s} \sqrt{\hat B_{B_s}})/(f_{B_d} \sqrt{\hat B_{B_d}})$, as obtained from the UT analysis including angle and $|V_{ub}|/|V_{cb}|$ constraints, and from lattice QCD calculations~\cite{Lubicz:2008am}.}
\label{tab:lattice}
\end{table}
%%%%%%%%%%%%%%%%%%%%%%%%%%%%%%%%%%%%%%%%%%%%%%%%%%%%%%%%%%%%%%%%%%
\section{Unitarity Triangle Analysis beyond the Standard Model}
\label{NP}
In this section we present the update of the NP UT analysis, that is the UT analysis generalized to include possible NP effects.
This analysis consists first in generalizing the relations among the experimental observables and the elements of the CKM matrix, introducing effective model-independent parameters that quantify the deviation of the experimental results from the SM expectations.
The possible NP effects considered in the analysis are those entering neutral meson mixing.
Thanks to recent experimental developments, in fact, these $\Delta F=2$ processes turn out to provide stringent constraints on possible NP contributions.
In the case of $B_{d,s}$-$\bar B_{d,s}$ mixing, a complex effective parameter is introduced, defined as
\begin{equation}
\qquad \qquad \qquad \qquad C_{B_{d,s}}\,e^{2 i \phi_{B_{d,s}}} = \frac{\langle B_{d,s} | H_{eff}^{full}| \bar B_{d,s} \rangle}{\langle B_{d,s} | H_{eff}^{SM}| \bar B_{d,s} \rangle}\,,
\end{equation}
being $H_{eff}^{SM}$ the SM $\Delta F=2$ effective Hamiltonian and $H_{eff}^{full}$ its extension in a general NP model, and with $C_{B_{d,s}}=1$ and $\phi_{B_{d,s}}=0$ within the SM.
All the mixing observables are then expressed as a function of these parameters and the SM ones (see refs.~\cite{Bona:2005eu,Bona:2006sa,Bona:2007vi} for details).
In a similar way, for the  $K$-$\bar K$ system, one can write
\begin{equation}
\qquad C_{\epsilon_K} = \frac{Im[\langle K | H_{eff}^{full}| \bar K \rangle]}{Im[\langle K | H_{eff}^{SM}| \bar K \rangle]}\,,\qquad \qquad
C_{\Delta m_K} = \frac{Re[\langle K | H_{eff}^{full}| \bar K \rangle]}{Re[\langle K | H_{eff}^{SM}| \bar K \rangle]}\,,
\end{equation}
with $C_{\epsilon_K}=C_{\Delta m_K}=1$ within the SM.
In order to take into account the large long-distance uncertainty in $\Delta m_K$, we include a long-distance contribution varying with a uniform distribution between zero and the experimental value of $\Delta m_K$.
For $D$-$\bar D$ mixing, the results of the analysis performed in ref.~\cite{Ciuchini:2007cw} are used.
On the other hand, the experimental constraints on the tree-level observables $\gamma$ and $|V_{ub}/V_{cb}|$, where NP contributions are expected to be negligible, are implemented as in the SM UT analysis, as well as the theoretical inputs.

\begin{figure}[t!]
\begin{center}
\vspace{-1.0cm}
\includegraphics[scale=0.3,angle=0]{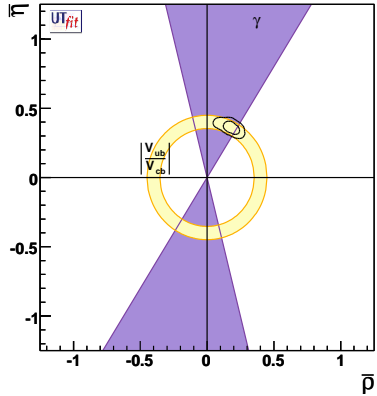} \\
\end{center}
%\vspace{-0.3cm}
\caption{\sl Constraints on the $\bar \rho$-$\bar \eta$ plane, in the NP UT analysis. The closed contours at $68$\% and $95$\% probability are shown. The full lines correspond to $95$\% probability regions for the different constraints.}
%\vspace{-0.2cm}
\label{fig:NP_allconstr}
\end{figure}

\begin{figure}[!t]
\vspace{-1.0cm}
\begin{minipage}{16pc}
\includegraphics[scale=0.3,angle=0]{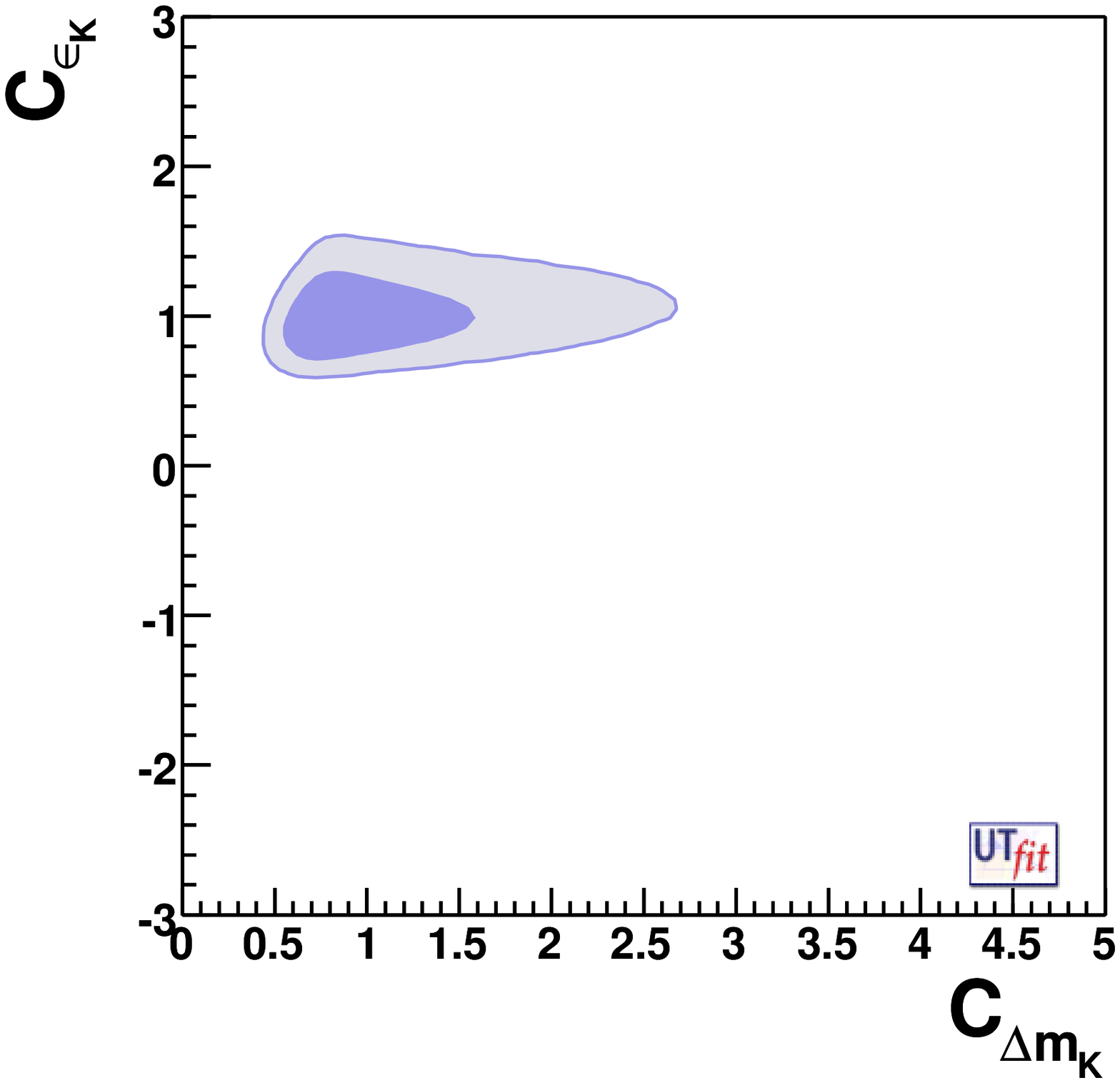}
\end{minipage}\hspace{2pc}
\begin{minipage}{16pc}
\includegraphics[scale=0.3,angle=0]{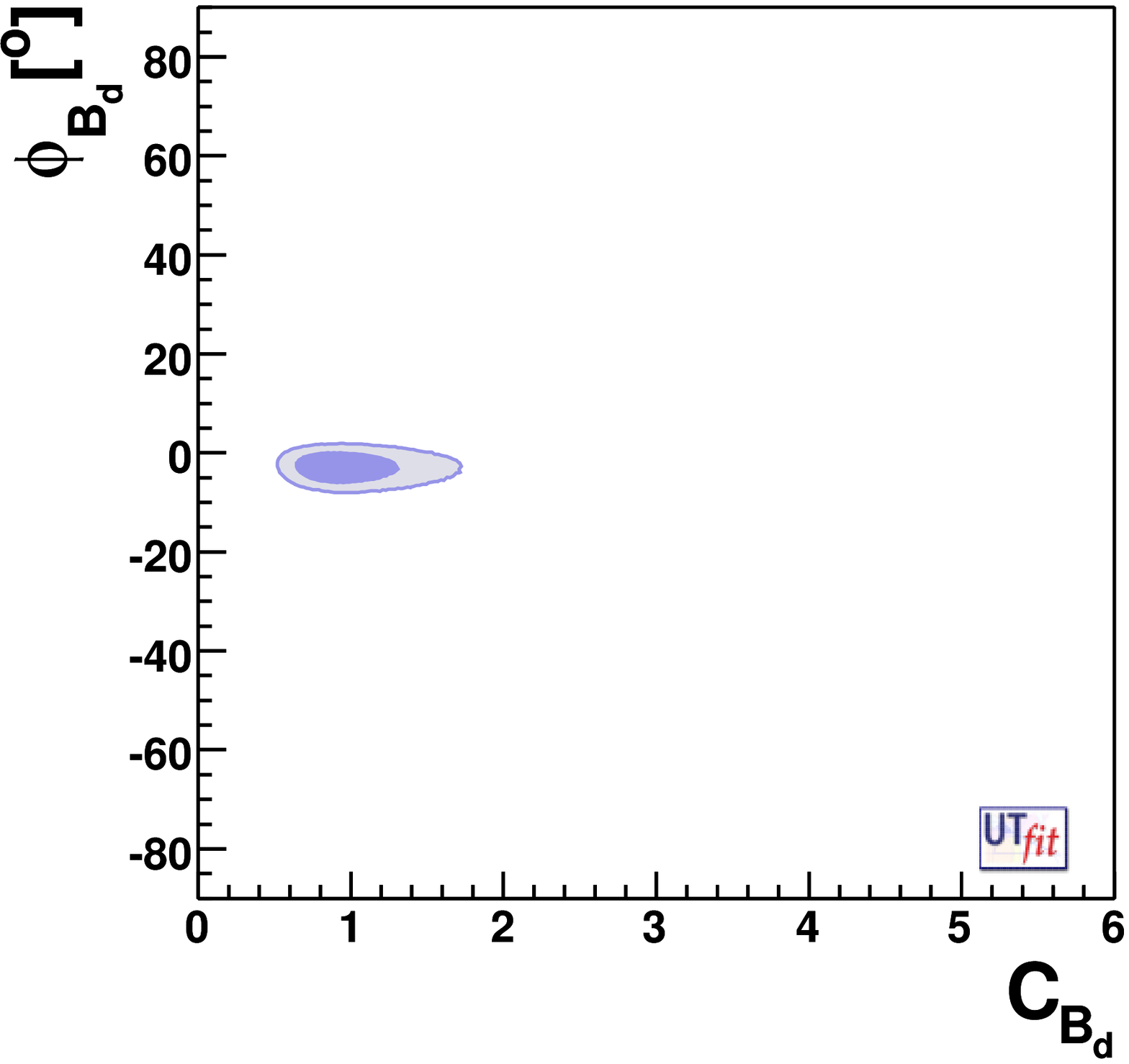}
\end{minipage}\hspace{2pc}
\begin{minipage}{16pc}
\includegraphics[scale=0.3,angle=0]{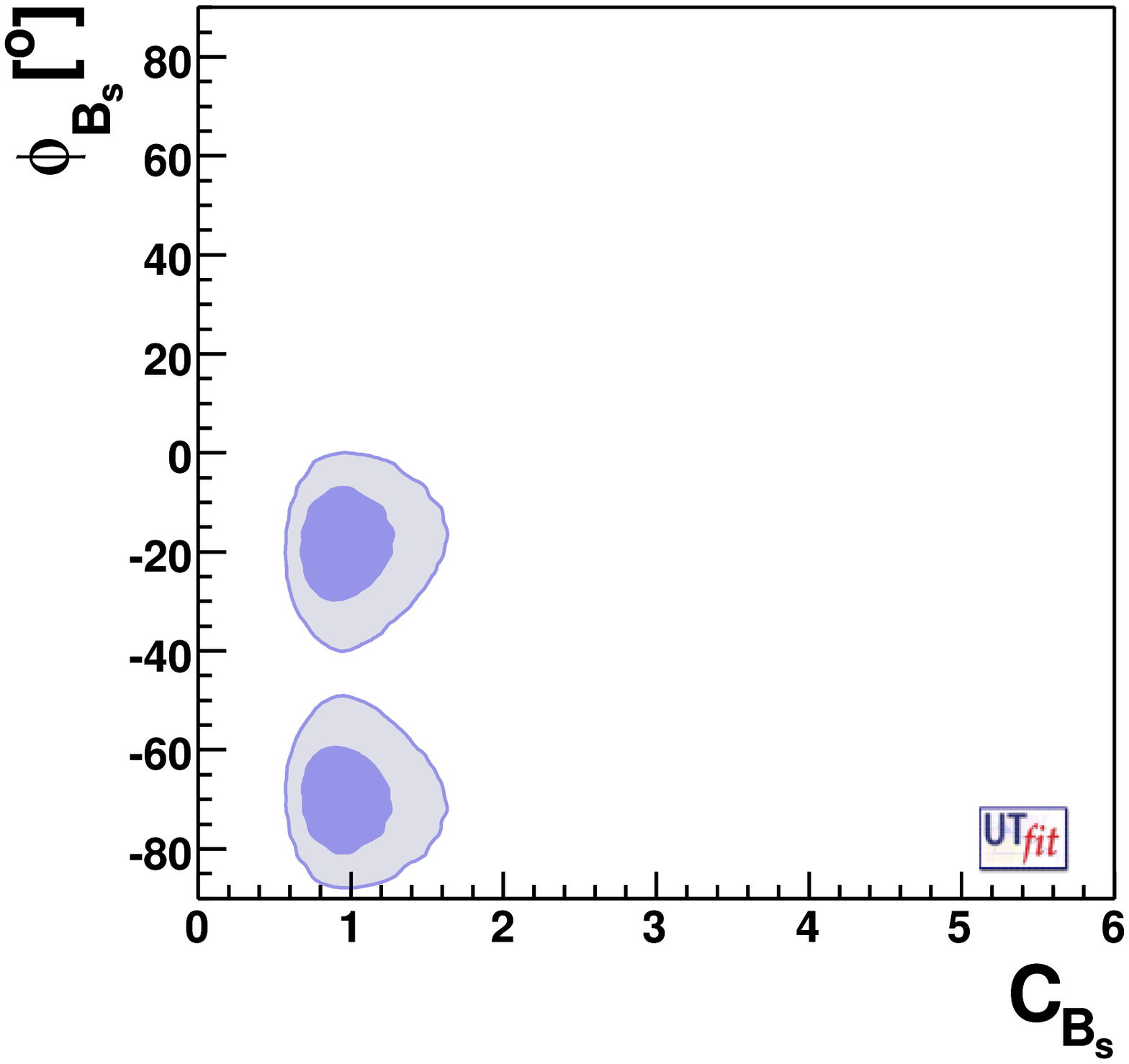}
\end{minipage}\hspace{2pc}
\begin{minipage}{16pc}
\includegraphics[scale=0.3,angle=0]{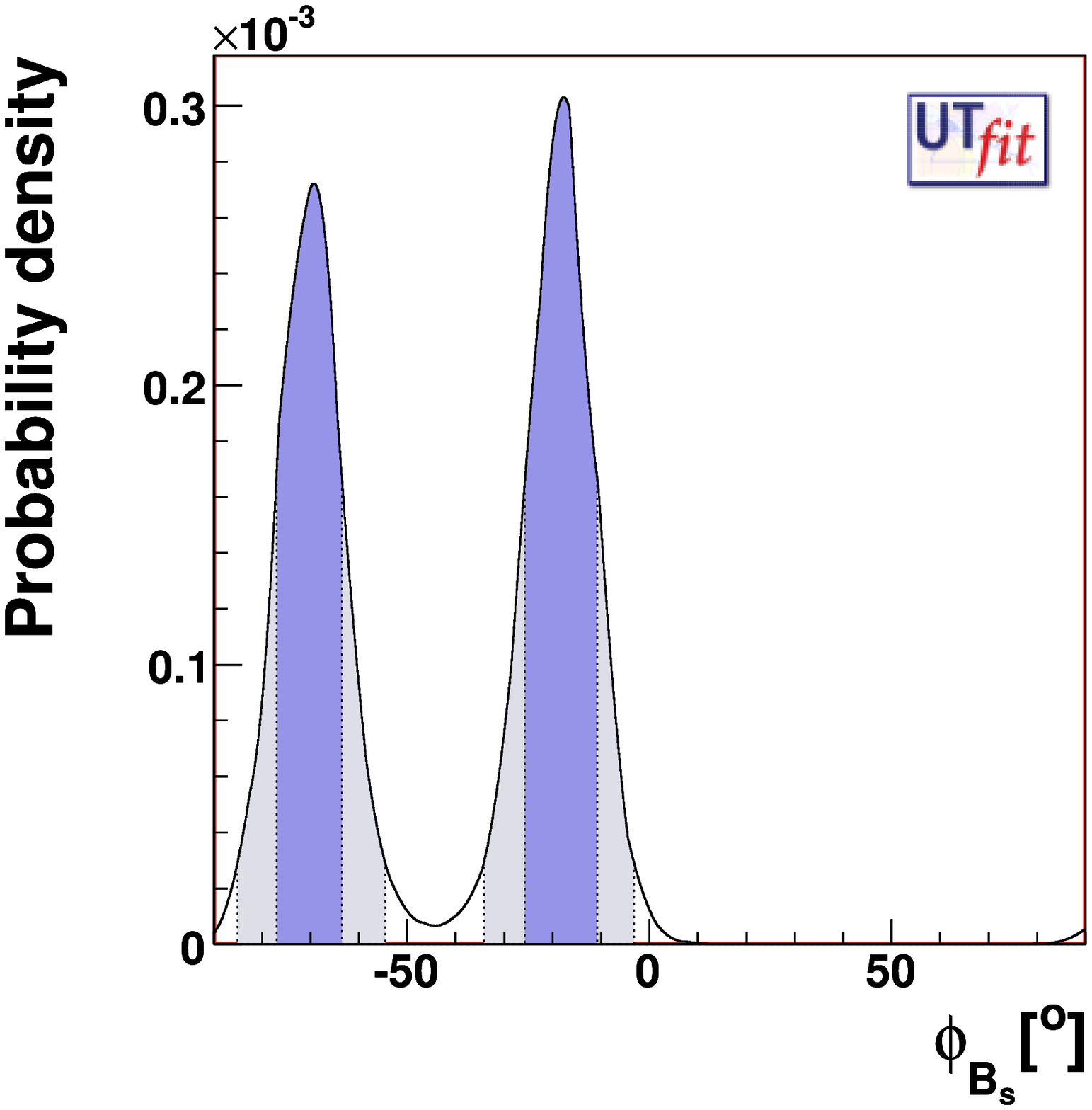}
\end{minipage}
\caption{\sl Determination of the NP parameters:  $68$\% and $95$\% probability regions for $C_{\epsilon_K}$ vs. $C_{\Delta_{m_K}}$ (top left), for $\phi_{B_d}$ vs. $C_{B_d}$ (top right) and for $\phi_{B_s}$ vs. $C_{B_s}$ (bottom left), and probability density function for $\phi_{B_s}$.} 
\label{fig:NPpar}
\end{figure}

In this way, the combined fit of all the experimental observables selects the region of the $(\bar \rho, \bar \eta)$ plane shown in fig.~\ref{fig:NP_allconstr} ($\bar \rho=0.177\pm0.044$, $\bar \eta=0.360\pm0.031$), which is consistent with the results of the SM analysis.
This indicates that NP can only show up as a small correction to the SM CKM picture.
The fit also constraints the effective NP parameters, as shown in fig.~\ref{fig:NPpar}.
%%%%%%%%%%%%%%%%%%%%%%%%%%%%%%%%%%%%%%%%%%%%%%%%%%%%%%%%%%%%%%%%%%%%%
%%%%%%%%%%%%%%%%%%%%%%%%%%%%%%%%%%%%%%%%%%%%%%%%%%%%%%%%%%%%%%%%%%%%%

For $K$-$\bar K$ mixing~\footnote{The additional corrections to $\epsilon_K$ pointed out and estimated to be around $8$\% in ref.\cite{Buras:2008nn}, are not yet included in the analysis.}, the NP parameters are found in perfect agreement with the SM expectations, whereas the $B_d$-$\bar B_d$ mixing phase $\phi_{B_d}$ is found $\simeq 1.5 \sigma$ away from the SM expectation, reflecting the tension between the direct measurement of $\sin 2 \beta$ and its indirect determination from the other UT constraints, so that a further improvement of the experimental accuracy is looked forward.

The $B_s$-meson sector, where the tiny SM mixing phase $\sin 2 \beta_s \simeq 0.041(4)$ could be highly sensitive to a NP contribution, represents a privileged environment to search for NP.
In this sector, an important experimental progress has been achieved at the Tevatron collider in 2008 when both the CDF~\cite{Aaltonen:2007he} and D0~\cite{:2008fj} collaborations published the two-dimensional likelihood ratio for the width difference $\Delta \Gamma_s$ and the phase $\phi_s=2(\beta_s-\phi_{B_s})$, from the tagged time-dependent angular analysis of the decay $B_s \to J_{\psi} \phi$. 
Updating the UTfit analysis of ref.~\cite{Bona:2008jn}, by combining the CDF and D0 results including the now available D0 two-dimensional likelihood without assumptions on the strong phases, we find $\phi_{B_s}=(-69\pm7)^\circ \cup (-19\pm8)^\circ$, which is $2.9\sigma$ away from the SM expectation $\phi_{B_s}=0$.\footnote{The Heavy Flavour Averaging Group (HFAG)~\cite{HFAG}, treating $\Delta \Gamma_s$ and $\phi_s$ as independent quantities, finds a deviation of $\phi_{B_s}$ from the SM at  the level of $2.2 \sigma$.}
%%%%%%%%%%%%%%%%%%%%%%%%%%%%%%%%%%%%%%%%%%%%%%%%%%%%%%%%%%%%%%%%%%%%%
%%%%%%%%%%%%%%%%%%%%%%%%%%%%%%%%%%%%%%%%%%%%%%%%%%%%%%%%%%%%%%%%%%%%%
It will be interesting to see if this hint of NP will be confirmed once the Tevatron measurements will improve, in particular when the CDF collaboration will make the
new likelihood, based on an enlarged data sample of $2.8 fb^{-1}$, publicly available in a format that people can use.
We note that this NP signal would be not only a signal of physics beyond the SM but more in general beyond minimal flavour violation (MFV)~\cite{D'Ambrosio:2002ex,Buras:2000dm}, since a value of $\phi_{B_s}$ different from zero can only be an effect of a new source of flavour violation different from the Yukawa couplings.
Furthermore, this signal would hint a clear pattern of NP contributions to flavour violation, with strongly suppressed $s \leftrightarrow d$ transitions, small $b \leftrightarrow d$ transitions and visible $b \leftrightarrow s$ transitions, which could be explained by non-abelian flavour symmetries and in some supersymmetric theories of Grand Unification.
An alternative explanation to the observed size of $\phi_{B_s}$ has been recently given in ref.~\cite{Botella:2008qm}, in the framework of a model with violation of $3 \times 3$ unitarity.   

A further step in the NP analysis can be performed by adopting an approach based on the effective field theory, that is by describing the $\Delta F=2$ processes through an effective Hamiltonian that includes possible NP effects in additional operators with respect to the SM and in modifications of the Wilson coefficients.
The Wilson coefficients can be schematically written as $C_i(\Lambda) = (L\,F_i)/\Lambda^2$,
where $L$ distinguishes NP contributions originated by tree-level interactions ($L \sim 1$) from loop NP effects ($L \sim \alpha_s^2, \alpha_W^2$), $F_i$ represents the CKM factor in the SM and in MFV models while it differs form the CKM factor in the presence of new sources of flavour violation, and $\Lambda$ denotes the NP scale.
The experimental constraints on the Wilson coefficients can then be translated into bounds on the NP scale ($\Lambda=\sqrt{L\,F_i/C_i(\Lambda)}$), once some information for $L$ and $F_i$ is specified for the NP model.
As in ref.~\cite{Bona:2007vi} we consider here three main classes of NP models: i) MFV models, where there are no new operators in addition to the SM ones nor new sources of flavour violation; ii) next-to-MFV (NMFV) models, where again the only sources of flavour violation are the Yukawa couplings as in the SM, but new operators can be present; iii) the most general case of a NP model with new sources of flavour violation and operators.
By switching on one operator per time, the UT experimental constraints provide bounds on the Wilson coefficients and therefore on the NP scale, as shown in Table~\ref{tab:NPscale}.
We note that the data on $K$-$\bar K$ and $B_d$-$\bar B_d$ mixing, where at present no NP evidence is seen, provide a lower bound on the NP scale, whereas the recent hint of NP in $B_s$-$\bar B_s$ mixing yields an upper bound.
The comparison of lower and upper bounds shows that, if the NP signal in $B_s$-$\bar B_s$ mixing is confirmed, MFV and NMFV models are ruled out, and that NP has to show up with new sources of flavour violation with a hierarchical flavour pattern.
%%%%%%%%%%%%%%%%%%%%%%%%%%%%%%%%%%%%%%%%%%%%%%%%%%%%%%%%%%%%%%%%%%
\begin{table}[h!]
\begin{minipage}{16pc}
\begin{tabular}{||c||c|c|c||}
\hline 
LOWER BOUNDS & tree & $\alpha_s$ loop & $\alpha_W$ loop \\ \hline
MFV & $5.5$ & $0.5$ & $0.2$ \\
NMFV & $62$ & $6.2$ & $2$ \\
General & $24000$ & $2400$ & $800$ \\
\hline
\end{tabular}
\end{minipage}\hspace{2pc}
\begin{minipage}{16pc}
\begin{tabular}{||c||c|c|c||}
\hline 
UPPER BOUNDS & tree & $\alpha_s$ loop & $\alpha_W$ loop \\ \hline
NMFV & $35$ & $4$ & $2$ \\
General & $800$ & $80$ & $30$ \\
\hline
\end{tabular}
\end{minipage}
\vspace{0.2cm}
\caption{\sl The left table shows the lower bounds in TeV for the NP scale $\Lambda$ as obtained from the experimental constraints on $K$-$\bar K$ and $B_d$-$\bar B_d$ mixing. The right table shows instead the upper bounds in TeV provided by the experimental constraints on $B_s$-$\bar B_s$ mixing. The bounds refer to the three large classes of NP models explained in the text: MFV models, NMFV models and NP models with a general flavour structure. The three columns correspond to Wilson coefficients originated by tree-level NP contributions, strong loop interactions and electroweak loop interactions, respectively.}
\label{tab:NPscale}
\end{table}
%%%%%%%%%%%%%%%%%%%%%%%%%%%%%%%%%%%%%%%%%%%%%%%%%%%%%%%%%%%%%%%%%%

%
\end{document}